\documentclass[twocolumn,apj]{emulateapj} 
\usepackage{times}

\newcommand{\etal}{{et al.}}
\newcommand{\eg}{{\it e.g.,}}
\newcommand{\ie}{{\it i.e.,}}
\newcommand{\kms}{{km s$^{-1}$}}
\newcommand{\cc}{{cm$^{-3}$}}

\shorttitle{Extended Emission-Line Region of 3C 249.1}
\shortauthors{Fu \& Stockton}
\journalinfo{Accepted by ApJ}
\submitted{Accepted by ApJ}

\begin{document}
\title{
Integral Field Spectroscopy of the Extended Emission-Line Region of 3C 249.1\footnotemark[1] 
}

\footnotetext[1]{Based on observations obtained at the Gemini
Observatory, which is operated by the Association of Universities for
Research in Astronomy, Inc., under a cooperative agreement with the NSF
on behalf of the Gemini partnership: the National Science Foundation
(United States), the Particle Physics and Astronomy Research Council
(United Kingdom), the National Research Council (Canada), CONICYT
(Chile), the Australian Research Council (Australia), CNPq (Brazil) and
CONICET (Argentina).}

\author{Hai Fu and Alan Stockton}
\affil{Institute for Astronomy, University of Hawaii, 2680 Woodlawn
 Drive, Honolulu, HI 96822}

\begin{abstract}
We present Gemini Multiobject Spectrograph integral field spectroscopy
of the extended emission-line region associated with quasar 3C\,249.1.
The kinematics of the ionized gas measured from the [O\,{\sc iii}]\
$\lambda$5007 line is rather complex and cannot be explained globally by
a simple dynamical model, but some clouds can be modeled individually as
having locally linear velocity gradients. The temperatures of the
ionized gas appear uniform (varying from $\sim$12000 to 15000 K), while
the densities vary from a few tens to a few hundreds cm$^{-3}$. The
emission mechanism of all of the emission clouds, as indicated by the
line-ratio diagnostics, is consistent both with ``shock + precursor" and
pure photoionization models. The total mass of the ionized gas is on the
order of $10^9$ M$_\odot$. We estimate the bulk kinetic energy and
momentum of the extended emission-line region of $2.5\times10^{57}$ ergs
and $10^{50}$ dyne s, and a dynamical timescale of $\sim$10 Myr.  By
comparing the injection rates of kinetic energy and momentum of
different galactic wind models with the observation, we argue that the
emission-line clouds are most likely a direct result from the feedback
of the quasar.  We also discuss the nature of the extended X-ray
emission surrounding the quasar.   
 
\end{abstract}

\keywords{galaxies:interactions --- quasars: individual(3C 249.1)}

\section{Introduction}
About a third of low-redshift (z $\leq$ 0.45) steep-spectrum
radio-loud quasars are surrounded by luminous emission-line regions that
often extend to several tens of kpc from the nucleus \citep{sto87}.
The morphologies of these extended emission-line regions (EELRs) are often complex and
clearly display a non-equilibrium situation showing strings of knots and
filaments straddling tens of kpc, and they are in general uncorrelated
with either the distribution of stars in the host galaxy or the
structure of the extended radio source, if present. 

The origin of the gas comprising these EELRs remains uncertain. The
apparent correlation between the occurrence of EELRs and evidence for
strong interaction inspired the idea that the gas is tidally disrupted
material from a disk \citep{sto87}.
But this is unlikely
because (1) there is generally no correlation between stellar tidal
features and the distribution of the gas, and (2) some confinement
mechanism is required for the gas to retain the densities implied by the
emission-line spectra for more than a very brief time
($\sim10^4$--$10^6$ years; \citealt{fab87,sto02}).
It has also been suggested that the gas comes from a cooling flow in a
hot halo surrounding the quasar \citep{fab87}.
This scenario has
been ruled out by deep {\it Chandra} X-ray observations of four quasars
showing strong EELRs, since the hot (10$^8$ K) halo gas from which the
warm emission-line gas is suggested to condense is not detected,
indicating strong cooling is not taking place \citep{sto06}.
And, in any case, the standard cooling-flow paradigm seems to have been
ruled out by recent high-resolution X-ray spectroscopy \citep{pet03}.

If the gas cannot come from the intergalactic medium (IGM) through a
cooling flow, then it must originate from the interstellar medium (ISM)
of the quasar host galaxy. Despite the non-detection of hot gaseous
halos by the {\it Chandra} observations, highly structured X-ray
emission regions are seen around two of the four quasars surveyed. One
of the plausible explanations for the X-ray emission is that it is
thermal bremsstrahlung from high-speed shocks. On the other hand,
detailed photoionization modeling indicates that a low-density
phase ($\sim$2 \cc) and a high-density phase ($\sim$500 \cc) are needed
to explain the optical spectrum of one of the EELR clouds around
4C\,37.43 (E1; \citealt{sto02}). The two phases cannot be in
pressure equilibrium because their temperatures are almost the same. A
reasonable possibility is that the high densities in the gas are continuously 
regenerated by shocks propagating through the surrounding medium. 

The high-speed shocks could be driven by a galactic wind, which itself
results from the feedback from the quasar \citep{diM05} 
or a vigorous starburst in the host galaxy (see
\citealt{vei05} for a review). This evidence of
ongoing galactic winds combined with the morphological properties of
EELRs thus suggest a scenario where the EELRs are a direct result of a
superwind---the gas is originally the ISM in the host galaxy and has
been blown out and shocked by a galactic wind; the gas is then
photoionized by the UV continuum from the quasar and/or ionized by the
shocks. 

3C\,249.1 ($z = 0.31$) is a powerful quasar, also known as PG\,1100+772. 
Around the quasar, distinct X-ray emission
regions \citep{sto06} and one of the most luminous optical
EELRs at $z < 0.5$ \citep{sto87} are seen extending to
radial distances of tens of kpc, making it an ideal candidate for an
in-depth study.  In this paper, we take advantage of the simultaneous
spatial and spectral coverage of integral field spectroscopy (IFS) to
further explore the origin of the EELR and the extended X-ray emission
around 3C\,249.1. Throughout we assume a flat cosmological model with
$H_0=70$ km s$^{-1}$, $\Omega_m=0.3$, and $\Omega_{\Lambda}= 0.7$.


\section{Observations and data reduction\label{obs}} 
3C\,249.1 was observed with the Integral Field Unit (IFU) of the Gemini
Multiobject Spectrograph (GMOS) on the Gemini North telescope. The queue
observations were executed  in the early half nights of April 7 and 8,
2005 (UT). Three exposures of 2825 s were obtained with B600/G5303 grating at
a central wavelength of 6537 \AA. Between exposures the telescope was
offset by about 0\farcs4 to improve the spatial sampling. 
The half-field (one slit) mode was used to cover the emission lines from
[\ion{Ne}{5}]\,$\lambda3426$ to [\ion{O}{3}]\,$\lambda5007$.
With this setting, we had a field-of-view (FOV) of
$3\farcs5\times5\arcsec$, a wavelength range of 4400 to 7200 \AA, a
dispersion of 0.46 \AA\ pixel$^{-1}$ and an instrumental full width 
at half-maximum (FWHM) of 3.5
pixels (1.6 \AA). The seeing varied from 0\farcs8 to 1\farcs0.  

The data were reduced using the Gemini IRAF data reduction package
(Version 1.8). The data reduction pipeline (GFREDUCE) consists of the 
following standard steps:  bias subtraction, cosmic rays rejection, spectral extraction,
flat-fielding, wavelength calibration, sky subtraction and flux
calibration. Spectra from different exposures were assembled and
interpolated to construct individual datacubes (x,y,$\lambda$) with a
pixel size of 0\farcs05 (GFCUBE). Differential atmosphere refraction was then
corrected by shifting the image slices at each wavelength to keep the
location of the quasar constant. The datacubes were then binned to
0\farcs2 pixels, which is the original spatial sampling of the IFU
fiber-lenslet system.  

Since our study focuses on the emission line gas, it is desirable to
remove the light of the quasar from the datacubes. For 3C\,249.1 the
quasar component can be cleanly removed from the datacubes by assuming
that all of the continuum in the spectra is scattered light from the
quasar, because (1) previous studies have shown that the host galaxy of
3C\,249.1 is very faint in optical compared with the quasar itself, 
and (2) extended emission line clouds usually show essentially pure emission-line
spectra (\eg\ \citealt{sto02}). The detailed procedure was as
follows:  The model quasar spectrum was derived by cubic-spline
smoothing the average spectrum extracted from an aperture centered on
the quasar. Then the ratio of the spectrum from each pixel to this
template quasar spectrum was fit with a smooth curve comprising a small
number of cubic spline pieces, and the continuum was removed by
subtracting the quasar template normalized by this curve.  For each
spectrum, pixels near strong emission lines were excluded from the
sample of the continuum fitting, and the locations of these spectral
regions were determined by measuring the wavelength of the [O\,{\sc
iii}]\ $\lambda$5007 line from each residual spectrum. These steps were
repeated a few times to achieve best residual datacubes. Finally, the
residual datacubes were smoothed to a common spatial resolution, and
merged to a final datacube. We generated exposure maps for each
wavelength to account for the exposure time variation caused by the
dithering and the differential atmosphere refraction.               

\section{Results}

\subsection{Kinematics\label{kinematics}}
\begin{figure*}[!tb]
\epsscale{1.0}
\plotone{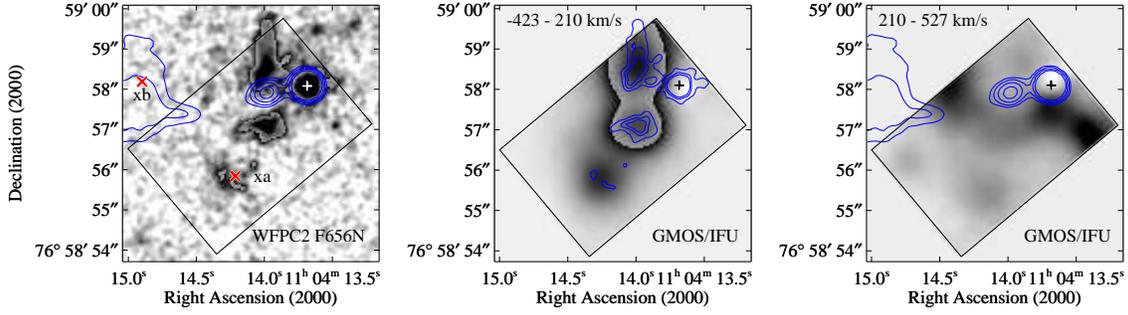}
\caption{{\it Left}---archived {\it HST}/WFPC2 F656N image of
3C\,249.1 (the redshifted [O\,{\sc iii}]\ $\lambda$5007 falls into this
filter), overlaid with contours of a VLA image at 5 GHz and 0\farcs35
resolution \citep{gil04}. Contours at $0.5\times(1,2,4,8)$ mJy
beam$^{-1}$.  The peak of the radio continuum has been registered to the
position of the quasar.  The two X-ray emission regions (xa \& xb) are
marked in red and labeled. The gray-scale images in this and the middle panel 
have been allowed to wrap around to show both low-surface-brightness detail and 
high-surface-brightness peaks. {\it Middle \& Right }---GMOS/IFU [O\,{\sc iii}]\
$\lambda$5007 radial velocity 
broad-band channel maps with contours of the {\it HST} image and contours
of the radio jet overlaid, respectively.  The radial velocity range 
(relative to that of the nuclear narrow
line region, $z = 0.3117$; negative velocities are blue shifted) is
shown in each panel.  
The rectangles delineate the $3\farcs5 \times 5\arcsec$ FOV of GMOS/IFU. 
The crosses indicate the position of the quasar, which has
been removed from the datacube (for details of the PSF removal, see \S~\ref{obs}).}
\label{fig:hstvla} \end{figure*}

\begin{figure*}[!tb]
\epsscale{1.0}
\plotone{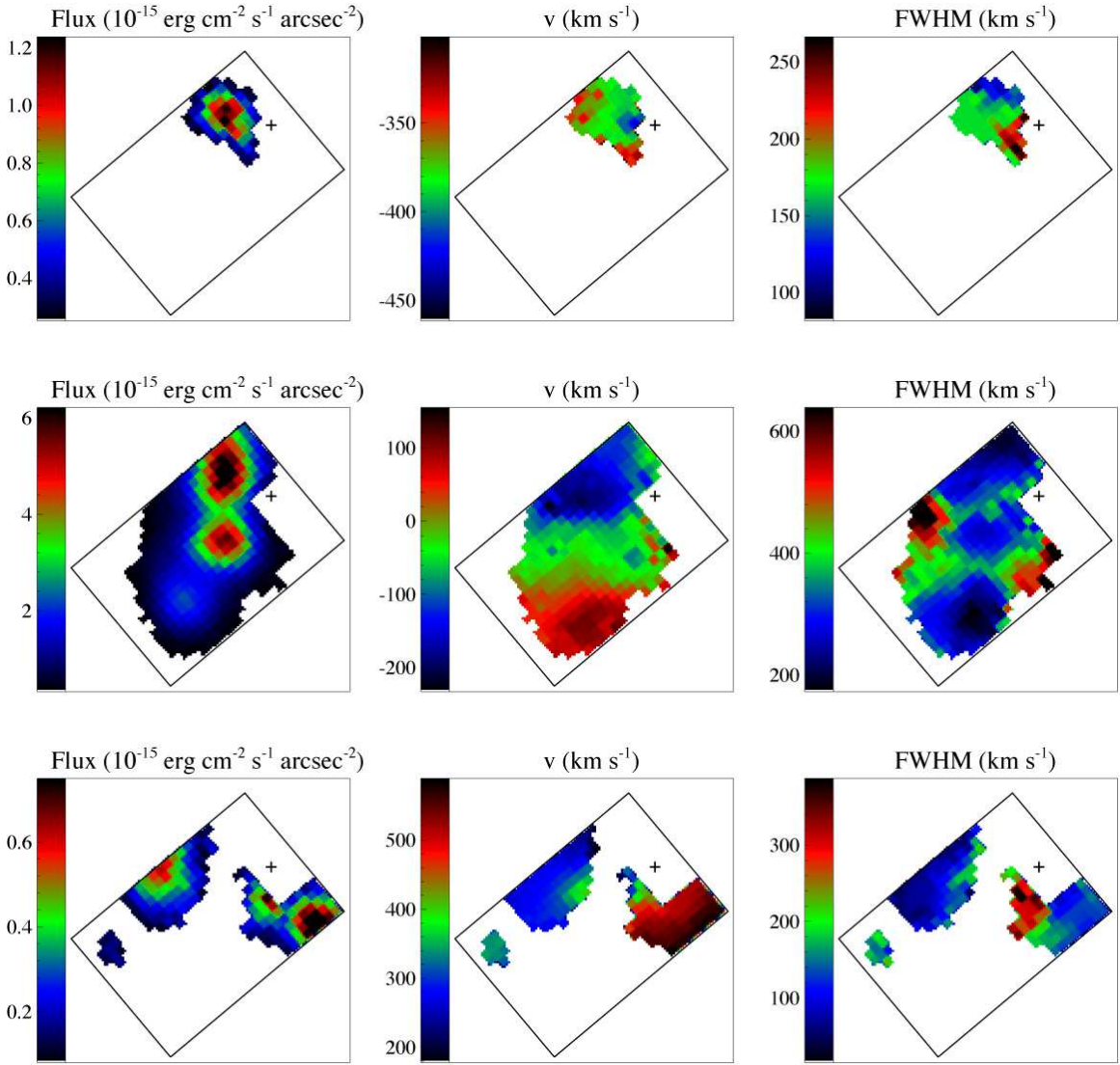}
\caption{Velocity field of 3C\,249.1 EELR derived from the
[O\,{\sc iii}]\ $\lambda$5007 region in the GMOS/IFU datacube. The three
columns, from left to right, are line intensity, radial velocity
(relative to that of the nuclear narrow line region) and velocity width
maps. 
To separate different clouds that are present
along the same line of sight, the velocity field is shown in three separate
velocity bands ($-$460 to $-$300 \kms, $-$230 to +150 \kms\ and +180 to +590 \kms\ 
from the top row to the bottom row). 
The rectangles delineate the $3\farcs5 \times 5\arcsec$ FOV of
GMOS/IFU. Pixels are 0\farcs2 squares. 
The crosses indicate the position of the quasar, which has been
removed from the datacube. For details see \S~\ref{kinematics}.} \label{fig:velofield}
\end{figure*}

Kinematics of the ionized gas can be measured from strong emission
lines. Since a single 47 min exposure is enough to acquire good S/N in
the [O\,{\sc iii}]\ $\lambda$5007 line region, we derived the velocity
fields from the single datacube that showed the best spatial resolution.
Figure~\ref{fig:hstvla} compares the broad-band images created from the
datacube and the {\it HST} WFPC2 image along with the radio jet.   The
velocity field is shown in Figure~\ref{fig:velofield} in three separate
velocity bands from $-$460 to +590 \kms\ (relative to the quasar
narrow-line emission) to separate different clouds that are present
along the same line of sight.  The FWHM measurements were corrected for
the $\sim$70 \kms\ instrumental resolution.  Thanks to the high spectral
resolution of the data set, at least eight EELR clouds at different velocities
are identified from the datacube. In the {\it HST} image, two of the
EELR clouds are blended together since they project onto each other, and many
are seen as faint knots because of the shallow exposure (300 s) and
the truncation at about +375 \kms\ of the F656N filter. 

The EELR exhibits rather complex global kinematics which cannot be
explained by a simple dynamical model. The three most luminous EELR clouds in
the velocity range of $-$230 to +150 \kms\ show a velocity gradient from
N to S that aligns with the clouds. This gradient may be an artifact of a common
disc rotation inherited from the host galaxy; but the high velocity
clouds at $v > 200$ \kms\ show a velocity gradient from E to W, which is
almost perpendicular to the previous one. Also, the southern cloud in
the $-$230 to +150 \kms\ panels shows a velocity increase of $\sim$100
km s$^{-1}$ along NE-SW direction, which is consistent with the
elongation observed in the {\it HST} image, thus it can be ascribed to a
simple rotating disk (at another different orientation); however, the
line widths also decrease along the same direction, suggesting two
discrete clouds are present as hinted in the {\it HST} image.  So
overall the kinematics are disordered, but on the other hand the
velocity field seems locally ordered. 

There are apparently two different species of emission-line clouds---one 
showing relatively broad lines (FWHM $\sim$300 \kms), the other
showing narrower lines (FWHM $\sim$100 \kms). The relatively broad line
regions are more luminous, have smaller radial velocities ($|v| < 200$
\kms), and are identified as spatially resolved blobs in the {\it HST}
image. In contrast, the narrower line regions tend to have larger radial
velocities ($|v| > 300$ \kms) and are faint and seen as unresolved knots
in the {\it HST} image. These knots feature an H$\beta$ luminosity of
$\gtrsim10^{40}$ erg s$^{-1}$. The most straightforward interpretation
is that the narrower line regions are less massive clouds, therefore
less luminous and more easily accelerated to a higher velocity; they
are reminiscent of the ``bullets" frequently seen in Herbig-Haro objects
and local galaxies exhibiting superwinds (\eg\ \citealt{cec01}) 
but are orders of magnitudes more energetic.

We notice that in the $-$230 to +150 \kms\ panels there is a region of
very low surface-brightness emission (east of the bright emission cloud
close to the center), but showing very broad lines (FWHM
$> 600$ \kms).  It cannot simply be a result of blended emission lines
scattered from the bright clouds nearby, since from the regions between
the clouds (refer to the middle right panel of Fig.~\ref{fig:velofield})
we can tell that blending can increase the line width up to only $\sim$400
\kms. The fact that the region is along the radio jet (see Fig.~\ref{fig:hstvla}) 
hints that the gas is disturbed by the radio jet. 

\subsection{Electron density and temperature}
\begin{figure*}[!tb]
\epsscale{1.0}
\plotone{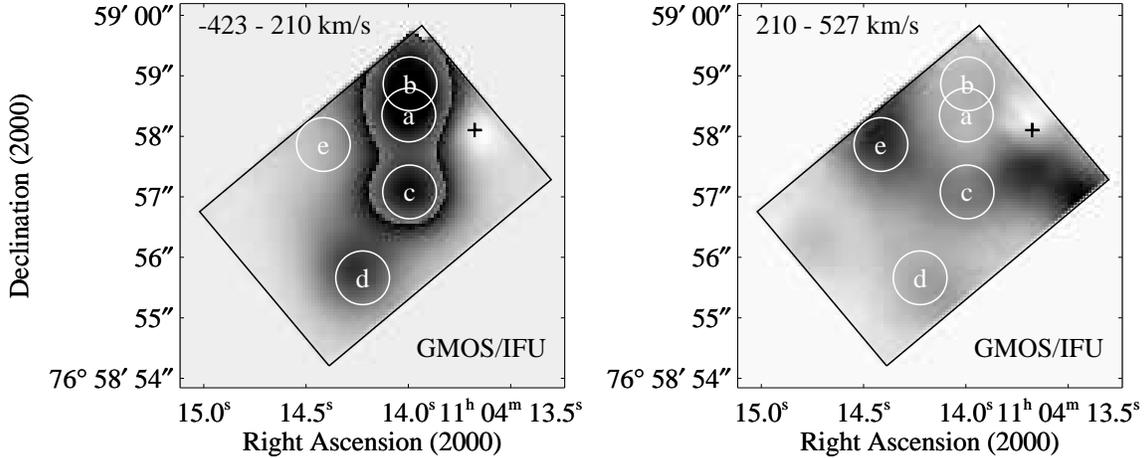}
\caption{The extraction apertures of different emission-line regions.
Background images are the GMOS/IFU [O\,{\sc iii}]\ $\lambda$5007 radial
velocity broad-band channel maps.  The FOV is slightly different
from the one shown in Figs.~\ref{fig:hstvla}\&\ref{fig:velofield},
because this one is created from the final merged datacube, while the
latter is from the single exposure which had the best spatial resolution.
} \label{fig:ex_regs}
\end{figure*}

\begin{figure}[!t]
\epsscale{1.25}
\plotone{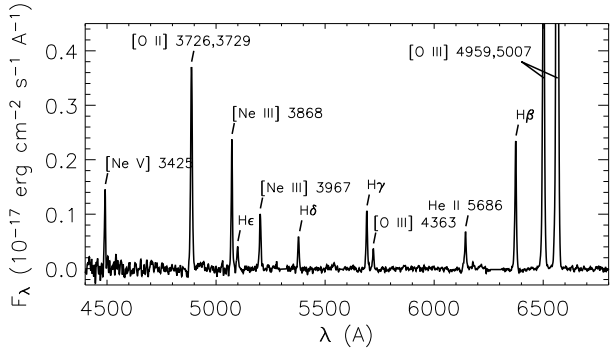}
\caption{Extracted spectrum of 3C\,249.1 EELR-$b$ (refer to Fig.~\ref{fig:ex_regs}), with important emission lines labeled.  }
\label{fig:labeledspec} \end{figure}

\begin{figure}[!t]
\epsscale{0.8}
\plotone{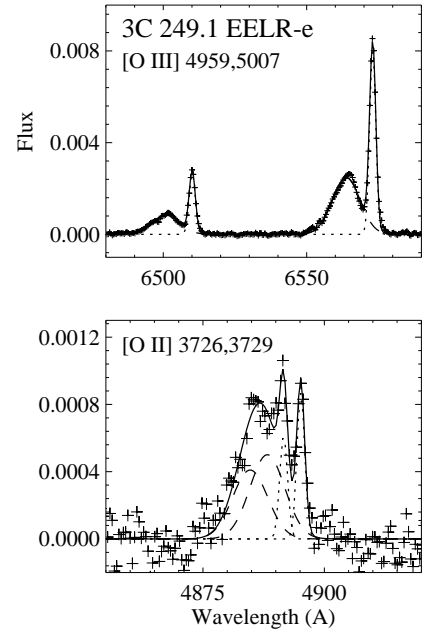}
\caption{ Line profiles of the [O\,{\sc iii}]\ $\lambda\lambda$4959,
5007 lines (upper panel) and the [O\,{\sc ii}]\ $\lambda\lambda$3726,
3729 doublet.  The spectrum is extracted from 3C\,249.1 EELR-e (see
Fig.~\ref{fig:ex_regs}). Note that two velocity components are present 
(the relatively broader component is a result of both the very broad-line 
cloud discussed in \S 3.1 and the contamination from the much brighter 
emission clouds nearby). The solid curves show the
best fit of four-Gaussian models, and the dashed curves and dotted
curves show the ``broad" and ``narrow" component, respectively. The fit
to the [O\,{\sc ii}] doublet have been constrained so that the relative
velocity between the ``broad" and ``narrow" components are frozen to
that determined from the fit to the [O\,{\sc iii}] lines.    }
\label{fig:rouT} \end{figure}

The luminosity-weighted average electron density can be determined from
the ratio of the [O\,{\sc ii}]\ $\lambda\lambda$3726, 3729 doublet, and
the electron temperature can be determined from the [O\,{\sc iii}]\
$\lambda$4363/($\lambda$4959+$\lambda$5007) intensity ratio (\eg\
Osterbrock 1989). The S/N ratio from a single 0\farcs2 pixel is not high
enough to perform this kind of measurement, so we combined spectra from
individual emission regions using an aperture of 0\farcs4 in radius and
a centered PSF as a weighting function to achieve optimal extraction. As
the last step, we correct the spectra for the line-of-sight Galactic
extinction (A$_V$ = 0.112; Schlegel \etal\ 1998), and for the intrinsic
reddening by dust associated with the clouds, which is estimated by
matching the measured H$\gamma$/H$\beta$ ratio with the value
predicted by case-B recombination. In both cases we used the standard
Galactic extinction curve of \citet{car89}.  Unlike
the kinematic analysis, these line ratios were extracted from the merged datacube in
order to obtain better S/N for faint lines. Figure~\ref{fig:ex_regs}
shows the extraction regions, and Fig.~\ref{fig:labeledspec} shows the
extracted spectrum of EELR-$b$ as an example. The spectra usually show
either asymmetric line profiles or two components with different radial
velocities. For both cases the line profiles usually can be fit quite
well with two Gaussians.
Figure~\ref{fig:rouT} shows an example of our two-Gaussian model fit to the line
profiles.  The spectrum shows two components at different radial
velocities.  The relatively broader component reflects both the very
broad-line cloud (\S \ref{kinematics}) that happens to lie in the same region and the
contamination from the much brighter emission clouds nearby
(EELR-$a,b,c$). We measure the velocities and FWHMs of the two
components from the [O\,{\sc iii}]\ $\lambda\lambda$4959,5007 lines
where we have the best S/N, and freeze these parameters when fitting the
much fainter [O\,{\sc iii}]\ $\lambda$4363 line and the [O\,{\sc ii}]
doublet. We then use the measured line fluxes to derive the electron
temperatures and densities for individual clouds.

The electron temperatures of different emission regions appear fairly
uniform, varying from $\sim$12000 to 15000 K. We also obtained good
measurements of electron densities from the relatively isolated clouds
(EELR-$b$, $d$ \& $e$), and we found that the density varies from a few
tens to a few hundreds cm$^{-3}$. Lower density gas has higher
ionization states (as indicated by the [O\,{\sc iii}]/[O\,{\sc ii}]
intensity ratios). The results are tabulated in Table 1.  Clearly the
pressure is quite different in different EELR clouds. 

\subsection{Constraints on the ionization mechanism}
Recent {\it Chandra} ACIS observations show that 3C\,249.1 has extended,
highly structured X-ray emission regions, one of which follows the
structure of the inner optical emission quite closely ($xa$ in
Fig.~\ref{fig:hstvla} {\it Left}; region $a$ in Fig.~4 of Stockton
\etal\ 2006). The X-ray emission of $xa$ cannot be connected with X-ray jets
frequently seen associated with radio-loud quasars, nor can it be due to
electron scattering of nuclear emission. It could either be thermal
bremsstrahlung emission from high-speed shocks {\it or} X-ray
recombination lines from the $10^4$ K photoionized gas. 
If it were the former case, one would expect shock-ionized gas near the
X-ray emitting region.

\begin{figure*}[!tb]
\epsscale{1.0}
\plotone{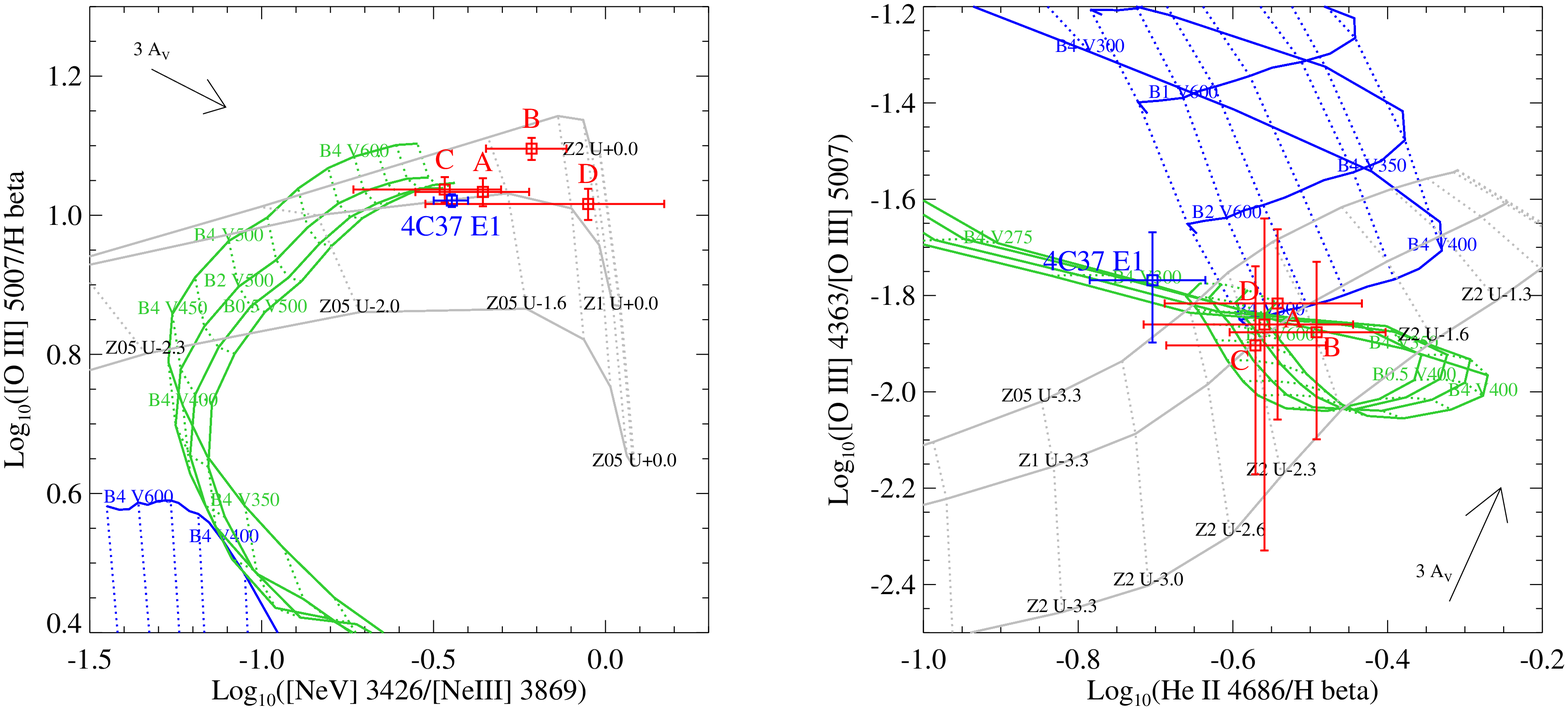}
\caption{
Optical line-ratio diagrams are overplotted with model grids. The
dusty radiation pressure-dominated photoionization model (Groves \etal\
2004), shock-only and ``shock + precursor" model (Dopita \& Sutherland
1996) grids are plotted in grey, blue and green, respectively.  The
photoionization grids assume a power-law photoionizing continuum with
index $\alpha = -1.4$ and show a range of ionization parameters
(Log$_{10}$(U)) for metallicities of 0.5 $Z_\odot$, 1.0 $Z_\odot$ and
2.0 $Z_\odot$.  The shock + precursor models assume equal contribution
to the H$\beta$ line from the shocked gas and the precursor. A range of
shock velocities ($V_S$ [km s$^{-1}$]) and magnetic parameters
($B/n^{1/2}$ [$\mu$ G cm$^{-3/2}$] ) are covered. Line ratio
measurements from four major EELR clouds of 3C\,249.1 are shown in red, and
the blue point shows the measurement from the Keck II/LRIS long-slit
spectrum of 4C\,37.43 E1 \citep{sto02}.
Arrows are reddening vectors of 3 A$_V$, assuming
the standard Galactic reddening law and R$_V$ = 3.1.}
\label{fig:lineratio} \end{figure*}

Because of the different shapes of the ionizing-photon spectrum and
different temperature regimes, ionization mechanisms (\eg\ shock,
``shock + precursor" and pure photoionization) can be distinguished
using diagnostic diagrams involving line ratios of some optical lines.
We measured line fluxes from the extracted spectra of various EELR clouds.
Figure~\ref{fig:lineratio} compares the observations with different model
grids.  Unfortunately, 3C\,249.1 EELR clouds (including the region associated
with X-rays---EELR-$d$) all fall in the regions where the
photoionization model and the ``shock + precursor" model overlap, making
it hard to distinguish the two, though the pure shock model can be ruled
out. We will discuss the X-ray emission regions in more detail in \S 4.1.

We notice that the 3C\,249.1 EELR shows systematically more He\,{\sc ii}
emission and lower [O\,{\sc iii}]\
$\lambda$4363/($\lambda$4959+$\lambda$5007) ratios than 4C\,37.43 E1,
indicating a higher metallicity and a lower temperature. But aside from
those lines, the spectra look remarkably similar to that of 4C\,37.43
E1. The measured line fluxes are tabulated in Table~2, along with
predictions from the two-phase photoionization models of \citet{sto02}.

\subsection{Mass of the ionized gas}

The mass of the ionized clouds can be derived from their H$\beta$
luminosity and electron density. For the EELR of 3C~249.1 within the
IFU field-of-view, we estimated a total H$\beta$ flux of
$2.9\times10^{-15}$ erg cm$^{-2}$ s$^{-1}$ 
by assuming a constant [O\,{\sc iii}]\ $\lambda$5007/H$\beta$ ratio of
11 (Fig.~\ref{fig:lineratio}). 
The mass of the \ion{H}{2} region is then 
$$ 
M_{\rm H} = \frac{4\pi m_p f_{\rm H\beta} d_L^2}{\alpha_{\rm H\beta} n_e h\nu}{\rm ,} 
$$
where $m_p$ is the proton mass, $d_L$ is the luminosity distance,
$\alpha_{\rm H\beta}$ is the effective recombination coefficient of
H$\beta$,  and $h\nu$ is the energy of a H$\beta$ photon \citep{ost89}.  
Assuming $n_e$ = 100 cm$^{-3}$ (\S 3.2) and case-B recombination
at 10$^4$ K, we obtain $M_{\rm H}$ = $6\times10^7$ M$_{\odot}$. A
similar calculation for 4C\,37.43 E1 yields a mass of $10^8$
M$_{\odot}$. 

These values are probably underestimates.
The mass derived above assumes a single phase cloud with an average
density $\sim$100 \cc. However, through detailed photoionization
modeling, \citet{sto02} found that at least two different
density regimes are required to reproduce the observed spectrum of
4C\,37.43 E1: a density-bounded component with $n_e \sim 2$ \cc and a
filling factor $f \sim 1$ and an ionization-bounded component with $n_e
\sim 500$ \cc and $f \sim 10^{-5}$, each contributing about one half of
the H$\beta$ flux. The spectra of the EELR clouds of 3C\,249.1 look remarkably
similar to that of 4C\,37.43 E1 (see Table~\ref{tab:lineratio}), so it
is appropriate to use the same modeling results. Since the [O\,{\sc ii}]
emission comes almost entirely from the high density component, the
luminosity-weighted electron density $\sim$100 \cc\ measured from
[O\,{\sc ii}]\ $\lambda\lambda$3726, 3729 doublet (\S 3.2) favors the
densest material. This density could be considerably larger than the
mass-weighted density (\ie\ weighting $\propto n_e f$) that should be
used in deriving the mass, therefore the mass derived above
($6\times10^7$ M$_{\odot}$) could have been seriously underestimated.
For example, assuming $n_{e1}$ = 2 \cc\ and $n_{e2}$ = 150 \cc\ and both
components contributing equally to the H$\beta$ luminosity, we obtain
$\sim$1.5$\times10^9$ M$_{\odot}$ of gas in the low density component and
only $\sim$2.0$\times10^7$ M$_{\odot}$ of gas in the high density
component. Note that this mass is only an order of magnitude lower than
the virial mass derived from assuming that the clouds are self gravitating and
that the measured line widths are entirely due to motions governed by
this gravitational potential (\ie\ ignoring turbulence or velocity gradients
along the line of sight): $M_{virial} = 5 R 
\sigma^2 / G \simeq 3.8 \times 10^{10} R_{1kpc}$ FWHM$_{300}^2$
M$_{\odot}$\footnote{1 kpc is the average radius of EELR-$c$ measured
from the {\it HST} image. The measurement was corrected for PSF
broadening.}. Considering the uncertainties in the simple two-phase
model (\eg\ the possibility of a large neutral fraction), the actual 
masses of individual clouds could potentially be
significantly larger than the masses derived from the line fluxes, and perhaps 
the clouds might even be gravitationally
bound. This could provide an alternative explanation for the existence
of high-density gas without invoking shocks. 

\subsection{Kinetic energy and momentum}
With the knowledge of the mass and kinematics of the emission-line
clouds, we can estimate their kinetic energy and momentum. The bulk
kinetic energy of a luminous EELR cloud is 
$E_{KE}$ = $M v^2 / 2$ = $2.5\times10^{57} M_9
v_{500}^2$ ergs, where $M_9$ is the mass of the ionized gas in units of
$10^9$ M$_{\odot}$ and $v_{500}$ is the velocity in units
of 500 \kms. The kinetic energy of the unresolved kinematic
substructures (``turbulent" kinetic energy) can be derived from the
measured line widths, $E_{turbulent} = 6.4\times10^{56} M_9$
FWHM$_{300}^2$ ergs. The emission-line gas also has a thermal energy of
$E_{TH} = 2.5 \times 10^{54} M_9 T_4$ ergs ($T_4 = T(K)/10^4 K$), which
is less than 0.1\% of the kinetic energy. Similarly, the total momentum
is $p_{kin}$ = $M v$ = $10^{50} M_9 v_{500}$ dyne s. 
Whatever the driving source of this outflow is, it must have deposited this enormous amount of energy into the EELR in a short period of time ($\sim$10 Myr; see \S~\ref{outflow}).
We will try to constrain the power source of this outflow in \S~\ref{outflow}.

\section{Discussion}

\subsection{The nature of the X-ray emission}
\begin{figure}[!tb]
\epsscale{1.2}
\plotone{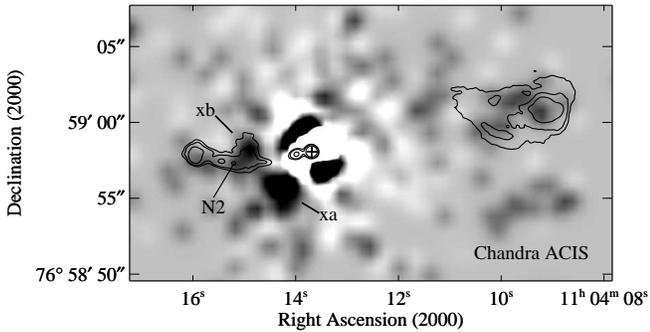}
\caption{PSF-subtracted {\it Chandra} ACIS image of 3C\,249.1 overlaid with contours of
a VLA image at 5 GHz and 0\farcs35 resolution \citep{gil04}.
Contours at $0.5\times(1,2,8)$ mJy beam$^{-1}$. The two X-ray emission regions (xa \& xb) 
and a radio hotspot (N2) are labeled. The peak of the radio continuum has been registered 
to that of the X-ray emission, which is indicated by a cross, before a scaled synthetic PSF 
was subtracted from the X-ray image (refer to \citealt{sto06} 
for more details on the PSF subtraction).} \label{fig:xrayvla} \end{figure}

In Fig.~\ref{fig:xrayvla} we compare the residual {\it Chandra} X-ray image of
3C\,249.1 and the 0\farcs35 resolution radio image. This figure shows 
convincingly that one of the X-ray emission region, $xb$, has an intimate
relationship with the radio jet. The X-ray emission is coincident with a
region protruding perpendicular to the jet direction, a feature that has
been noted by several authors (\eg\ designated X1 in \citealt{gil04}). 
Since there is a large offset ($\sim$1\farcs4 $\approx$ 6.4 kpc) between
$xb$ and the nearest hotspot (N2, following \citealt{gil04}), 
it might not simply be synchrotron radiation from the jet, but possibly a result of 
a gas cloud being shocked by the jet fluid.

The other X-ray emission region, $xa$, is covered by the IFU FOV.  Gas
kinematics, line-ratio diagnostics and its morphology all fail to
provide convincing evidence of shocks, so we can not rule out the
possibility that the X-ray emission is actually a result of the same
recombination process as what causes the optical emission. If the X-ray
is indeed from recombination lines, then the other high
surface-brightness clouds (EELR-$a,b,c$) should also be bright in X-ray; 
but it is impossible to detect the emission from the {\it Chandra} data, since
at such small radial distances their X-ray emission, if any, is completely 
overwhelmed by the X-rays from the quasar.

\subsection{Constraints on the power source of the outflow\label{outflow}}

The dynamical timescale of the EELR can be estimated from their radial
velocities and distances to the nuclei, $\tau_{dyn} \approx 10^7
D_{5kpc} \bar{v}_{500}^{-1}$ yr. Assuming an average velocity of 500
\kms\ is appropriate here, because the initial velocity of a galactic
wind is expected to be on the level of 1000 \kms\ and it has since been
slowed to the observed a few hundreds \kms\ due to mass entraining and
working against gravity. At the redshift of 3C\,249.1, the angular scale is
4.6 kpc arcsec$^{-1}$, so 5 kpc is about the projected distances of the
two brightest EELR clouds to the nucleus. The time scale is consistent with
the age of typical luminous radio lobes, $\sim10^7-10^8$ yr \citep{blu00}. 
The rate of mass outflow can be estimated assuming it is
constant over the dynamical age of the ionized gas,  $\dot{M} \simeq M /
\tau_{dyn} = 100 M_9 D_{5kpc}^{-1} \bar{v}_{500}$ M$_{\odot}$ yr$^{-1}$.
Similarly the injection rate of kinetic energy is $\dot{E}_{KE} \simeq
E_{KE} / \tau_{dyn} = 8\times10^{42} M_9 v_{500}^2 D_{5kpc}^{-1} \bar{v}_{500}$ ergs
s$^{-1}$, and the input momentum rate is $\dot{p}_{kin} \simeq p_{kin} /
\tau_{dyn} = 3\times10^{35} M_9 v_{500} D_{5kpc}^{-1} \bar{v}_{500}$ dyne. 
We emphasize that the rates can only be considered as lower limits to
the actual wind energetics since the majority mass of the wind flow may
not be detectable in the optical \citep{vei05} 
and a substantial amount of wind-entrained ISM may be shielded from the 
ionizing flux of the quasar.

If the outflow is driven by a starburst, a star formation rate (SFR) of
$\sim$10 M$_{\odot}$ yr$^{-1}$ is required to deposit enough kinetic
energy into the clouds, or $\sim$60 M$_{\odot}$ yr$^{-1}$ to inject
enough momentum into the ionized gas (Equations 2--3 in \citealt{vei05}. 
Since a significant fraction of the originally
injected kinetic energy may have been lost to radiation and
counteracting the gravitational potential, the kinetic energy is not
conserved. But the wind may still conserve momentum, so the SFR of
$\sim$60 M$_{\odot}$ yr$^{-1}$ is probably more realistic. The SFR
scales to an infra-red (IR) luminosity of $L_{IR}$(8-1000$\mu$m)
$\approx 5.8\times10^9 {\rm SFR} L_{\odot} = 3.5\times10^{11} L_{\odot}$
\citep{ken98}. Is this luminosity consistent with mid-IR
observations? 3C\,249.1 was detected in 12, 25 and 60 $\mu$m by IRAS
\citep{san89},  in 70 $\mu$m by MIPS \citep{shi05}, 
and in 100 $\mu$m by ISO \citep{haas03}.  
By using a model of two blackbody
components to fit the mid-IR spectral energy distribution (SED), we
obtain a total IR luminosity of $L_{IR}$(8-1000 $\mu$m) $\simeq
8.6\times 10^{11}$ L$_{\odot}$. 
Thus more than 40\% of the total IR luminosity must be powered by star
formation to explain the input momentum rate. However, the SED peaks
near 12 $\mu$m, so the IR emission is dominated by warm dust (T $\sim$
$100 - 200$ K) presumably directly heated by the quasar \citep{san89}.
From the SED fitting, less than 23\% of the total luminosity
could have come from a ``starburst" component with a characteristic dust
temperature $T \lesssim 65$ K. Apparently the IR data argues against an
on-going starburst as strong as $\sim$60 M$_{\odot}$ yr$^{-1}$, though
it doesn't formally rule out the possibility that there was an energetic
starburst which had caused the outflow but ceased not too long ago ($<
\tau_{dyn} \approx$ 10 Myr). 

Note also that the mass outflow rate is almost an order of magnitude larger than
the predicted mass injection rate from supernovae (SNe) for a
SFR$\sim$60 M$_{\odot}$ yr$^{-1}$ ($\dot{M}_{inj} \simeq 16$ M$_{\odot}$ yr$^{-1}$), which implies that the wind has
entrained considerable amount of gas from the ISM. This is in contrary
to the low efficiency for transporting mass out of the galaxy predicted by numerical
simulations of starburst-driven galactic winds \citep{str00}.

Now we consider the possibility that the outflow is directly driven by
the quasar of 3C\,249.1. Radiation from the quasar can couple to the
surrounding gas via various processes such as electric scattering,
photoionization, etc. The input momentum rate from radiation pressure is
$\dot{p} = 1.3\times10^{35} (\eta/0.1) \dot{M}_{acc}$ dyne s, where
$\eta$ is the radiative efficiency and $\dot{M}_{acc}$ is the mass
accretion rate of the black hole in units of M$_{\odot}$ yr$^{-1}$.
Hence a mass accretion rate of 2.5 M$_{\odot}$ yr$^{-1}$ is sufficient
to inject enough momentum to the clouds in $10^7$ yr. This accretion
rate is in agreement with the one implied by its bolometric luminosity,
$\dot{M}_{acc}$ = $8 (\eta/0.1)^{-1}$ M$_{\odot}$ yr$^{-1}$ (L$_{bol}$ =
$8\times10^{12}$ L$_{\odot}$; Sanders \etal\ 1989).

In summary, although we cannot totally exclude a starburst origin,
the available evidence favors the quasar as the power source of the
outflow.  If this is indeed the case, the presence of a luminous EELR
may tell us something about the quasar itself.  \citet{sto87} found
that roughly one-third to one-half of all powerful, steep-spectrum
radio-loud quasars show luminous EELRs, but there are clearly some
for which any extended emission is quite weak.  These latter cases
may be those for which the quasar ignition occurred sufficiently long
ago ($>$ a few $10^7$ years) that the expelled gas has dissipated to
the point that its surface brightness falls below our detection limit.

\acknowledgments
We thank Percy Gomez and Simon Chan for executing the queue observations, 
Julia Riley for providing the VLA images, 
and Lisa Kewley and Brent Groves for helpful discussions on the photoionization models. 
We thank the referee for a careful reading of the manuscript and for offering
suggestions that have led to a clearer presentation.
This research has been partially supported by NSF grant AST03-07335.

\clearpage

\begin{deluxetable}{lccccc}
\tablewidth{0pt}
\tablecaption{Electron Density and Temperature of the EELR Clouds.
\label{tab:rouT}}
\tablehead{
\colhead{} &\colhead{$V$}&\colhead{FWHM}&\colhead{$T_e$} &\colhead{$n_e$}&\colhead{} \\
\colhead{Region} &\colhead{(\kms)}&\colhead{(\kms)}&\colhead{(K)} &\colhead{(cm$^{-3}$)}&\colhead{[O\,{\sc iii}]/[O\,{\sc ii}]}}
\startdata
3C\,249 B&\phn$-$90&300&$12000\pm\phn800$&$160\pm100$&5.6\\ 
3C\,249 D&$+140$&300&$12100\pm1300$&$300\pm140$&4.2\\ 
3C\,249 E&$+300$&120&$15000\pm1500$&$\phn30\pm100$&7.2
\enddata
\end{deluxetable}

\begin{deluxetable}{lcccccc}
\tablewidth{0pt}
\tabletypesize{\footnotesize}
\tablecaption{Observed Line-Flux Ratios for 3C\,249.1 EELR Clouds and 4C\,37.43 E1 and Modeled Line-Flux Ratios
\label{tab:lineratio}}
\tablehead{
\colhead{Line Flux Ratios\tablenotemark{a}} & \colhead{[Ne\,{\sc v}]\,$\lambda3426$} & \colhead{[O\,{\sc ii}]\,$\lambda\lambda3726,9$\tablenotemark{b}}
& \colhead{[Ne\,{\sc iii}]\,$\lambda3869$} & \colhead{[O\,{\sc iii}]\,$\lambda4363$} &
\colhead{He\,{\sc ii}\,$\lambda4686$} & \colhead{[O\,{\sc iii}]\,$\lambda5007$}}
\startdata
 3C\,249 A     & $0.44\pm0.15$     & $2.39\pm0.17$     & $1.00\pm0.12$     & $0.16\pm0.07$     & $0.29\pm0.08$    & $10.80\pm0.50$\\
 3C\,249 B     & $0.60\pm0.15$     & $2.22\pm0.14$     & $0.98\pm0.11$     & $0.17\pm0.07$     & $0.32\pm0.07$    & $12.47\pm0.46$\\
 3C\,249 C     & $0.30\pm0.13$     & $2.52\pm0.19$     & $0.89\pm0.12$     & $0.14\pm0.06$     & $0.27\pm0.06$    & $10.89\pm0.46$\\
 3C\,249 D     & $0.63\pm0.40$     & $2.47\pm0.24$     & $0.71\pm0.13$     & $0.14\pm0.09$     & $0.28\pm0.08$    & $10.38\pm0.54$\\
 4C\,37 E1     & $0.34\pm0.04$     & $2.86\pm0.07$     & $0.95\pm0.05$     & $0.18\pm0.05$     & $0.20\pm0.03$    & $10.50\pm0.22$\\
\hline
    Model1\tablenotemark{c}              & 0.34              & 1.77              & 0.85              & 0.19              & 0.20             & 10.31 \\
    Model2              & 0.30              & 2.33              & 0.81              & 0.17              & 0.20              & 9.35
\enddata
\tablenotetext{a}{All line fluxes are given as ratios to the H$\beta$ flux}
\tablenotetext{b}{Total flux in the [\ion{O}{2}] doublet}
\tablenotetext{c}{Models 1 and 2 have 25\% and 33\% of the H$\beta$ flux coming from the 
500 \cc\ high-density component, see Stockton \etal\ 2002 (\S 3.4.2) for details}
\end{deluxetable}


\begin{thebibliography}


\bibitem[Blundell \& Rawlings(2000)]{blu00} Blundell, K.~M., 
\& Rawlings, S.\ 2000, \aj, 119, 1111

\bibitem[Cardelli et al.(1989)]{car89} Cardelli, J.~A., Clayton, G.~C.,
\& Mathis, J.~S.\ 1989, \apj, 345, 245 

\bibitem[Cecil et al.(2001)]{cec01} Cecil, G.,
Bland-Hawthorn, J., Veilleux, S., \& Filippenko, A.~V.\ 2001, \apj, 555,
338

\bibitem[Di Matteo et al.(2005)]{diM05} Di Matteo, T., 
Springel, V., \& Hernquist, L.\ 2005, \nat, 433, 604

\bibitem[Dopita \& Sutherland(1996)]{dop96} Dopita, M. A., \&
Sutherland, R. S.  1996, \apjs, 102, 161


\bibitem[Fabian et al.(1987)]{fab87} Fabian, A.~C., Crawford, C.~S.,
Johnstone, R.~M., \& Thomas, P.~A.\ 1987, MNRAS, 228, 963 

\bibitem[Gilbert et al.(2004)]{gil04} Gilbert, G.~M., Riley, J.~M.,
Hardcastle, M.~J., Croston, J.~H., Pooley, G.~G., \& Alexander, P.\
2004, \mnras, 351, 845 

\bibitem[Groves et al.(2004)]{gro04} Groves, B.~A., Dopita, M.~A., \&
Sutherland, R.~S.\ 2004, \apjs, 153, 9

\bibitem[Haas et al.(2003)]{haas03} Haas, M., et al.\ 2003, 
\aap, 402, 87

\bibitem[Kennicutt(1998)]{ken98} Kennicutt, R.~C.\ 1998, 
\apj, 498, 541

\bibitem[Osterbrock(1989)]{ost89} Osterbrock, D. E. 1989, Astrophysics
of Gaseous Nebulae and Active Galactic Nuclei, Mill Valley, California:
University Science Books

\bibitem[Peterson et al.(2003)]{pet03} Peterson, J.~R., Kahn, 
S.~M., Paerels, F.~B.~S., Kaastra, J.~S., Tamura, T., Bleeker, J.~A.~M., 
Ferrigno, C., \& Jernigan, J.~G.\ 2003, \apj, 590, 207

\bibitem[Sanders et al.(1989)]{san89} Sanders, D.~B., 
Phinney, E.~S., Neugebauer, G., Soifer, B.~T., \& Matthews, K.\ 1989, \apj, 
347, 29 

\bibitem[Schlegel et al.(1998)]{sch98} Schlegel, D.~J., Finkbeiner,
D.~P., \& Davis, M.\ 1998, \apj, 500, 525 

\bibitem[Shi et al.(2005)]{shi05} Shi, Y., et al.\ 2005,
\apj, 629, 88

\bibitem[Stockton et al.(2006)]{sto06} Stockton, A., Fu, H., Henry,
J.~P., \& Canalizo, G., 2006, \apj, 638, 635

\bibitem[Stockton \& MacKenty(1987)]{sto87} Stockton, A., \& MacKenty,
J. W. 1987, \apj, 316, 584

\bibitem[Stockton et al.(2002)]{sto02} Stockton, A., MacKenty, J. W.,
Hu, E. M., \& Kim, T.-S. 2002, \apj, 572, 735

\bibitem[Strickland \& Stevens(2000)]{str00} Strickland, 
D.~K., \& Stevens, I.~R.\ 2000, \mnras, 314, 511

\bibitem[Veilleux et al.(2005)]{vei05} Veilleux, S., Cecil, 
G., \& Bland-Hawthorn, J.\ 2005, \araa, 43, 769

\end{thebibliography}
\end{document}